\documentclass[preprint,aps]{revtex4}

\usepackage{xspace}
\usepackage{graphicx}
\usepackage{dcolumn}
\usepackage{bm}

\def\beq{\begin{equation}}
\def\eeq{\end{equation}}

\def\bea{\begin{eqnarray}}
\def\eea{\end{eqnarray}}

\newcommand{\roughly}[1]%
    {{\mathrel{\raise.3ex\hbox{$#1$\kern-.75em\lower1ex\hbox{$\sim$}}}}}
\newcommand{\lsim}{\mathrel{\roughly<}}

\newcommand{\scr}[1]{\ensuremath{\mathcal{#1}}}

\newcommand{\be}{\ensuremath{\beta}}
\newcommand{\ga}{\ensuremath{\gamma}}
\newcommand{\Ga}{\ensuremath{\Gamma}}

\newcommand{\De}{\ensuremath{\Delta}}
\newcommand{\ep}{\ensuremath{\epsilon}}

\newcommand{\ka}{\ensuremath{\kappa}}
\newcommand{\la}{\ensuremath{\lambda}}
\newcommand{\La}{\ensuremath{\Lambda}}

\newcommand{\GeV}{\ensuremath{\mathrm{~GeV}}}

\newcommand{\Eq}[1]{Eq.~(\ref{#1})}

\newcommand{\Ref}[1]{Ref.~\cite{#1}}

\begin{document}


\title{Natural $h \to 4g$ in Supersymmetric Models\\
and $R$-Hadrons at the LHC}

 \author{Markus A. Luty}\email{luty@physics.ucdavis.edu}
 \author{Daniel J. Phalen}\email{phalen@physics.ucdavis.edu}
\affiliation{%
Physics Department, University of California Davis,
Davis, CA 95616}%

\author{Aaron Pierce}\email{atpierce@umich.edu}
\affiliation{Michigan Center for Theoretical Physics, Department of Physics,
University of Michigan, Ann Arbor, MI 48109}

\date{\today}

\begin{abstract}
We construct a simple and natural supersymmetric model
where the dominant Higgs decay is $h \to aa$
followed by $a \to gg$.
In this case $m_h < m_Z$ is compatible with all experimental searches,
completely eliminating the fine tuning otherwise
required to satisfy Higgs search limits.
The model extends the MSSM with singlet Higgs fields as well as
vector-like colored particles that mediate the decay $a \to gg$.
The $a$ is a pseudo-Nambu Goldstone boson of a 
new global $U(1)$ symmetry, and can naturally have
any mass from a few GeV to $m_h/2$.
All interactions can be perturbative up to the GUT scale,
and gauge coupling unification is preserved if
the colored mediators come in complete GUT representations.
In this case $a \to \ga\ga$ has a $\sim 1\%$
branching ratio, so $h \to gg\ga\ga$ may be observable.
The colored particles that mediate the $a \to gg$ decay must be
below the TeV scale, and can therefore be produced at the LHC.
If these particles are stable on collider timescales, they will appear
as $R$-hadrons, a signal visible in early LHC running.
A smoking-gun signal that the 
stable colored particles are mediators of $h \to 4j$
is $R$-hadron production in association with an $a$.
We show that this signal with $a \to \ga\ga$
is observable at LHC with as little as 10 fb$^{-1}$ of
integrated luminosity.
Observation of $R$-hadrons plus missing energy would show that
the superpartner of the $R$-hadron is $R$-parity odd,
and therefore not an ordinary quark or gluon.
\end{abstract}


\maketitle

\section{Introduction}
Supersymmetry (SUSY) is a compelling framework for addressing the large
hierarchy between the weak scale and Planck scale.
Although there is no direct experimental evidence for the existence
of superpartners, there are several indirect indications that
SUSY is correct.
First, the minimal supersymmetric standard model (MSSM)
automatically predicts precise gauge coupling unification \cite{DRW}.
Second, the Higgs boson is naturally light in SUSY, so precision electroweak
constraints are automatically satisfied.
However, the Higgs boson is generically \emph{too} light in the MSSM:
at tree level $m_h \le m_Z$, while LEP searches
give a bound $m_h > 114\GeV$. Loop corrections can increase the Higgs boson mass only at the price
of fine tuning.
Since the Higgs vacuum expectation value (VEV) is fixed, the physical Higgs mass can be
increased only by increasing the quartic coupling.
In the MSSM this is accomplished by a heavy stop mass, which
gives a contribution to the quartic coupling of order
\beq
\De \la \sim \frac{N_c y_t^4}{16\pi^2} \ln\frac{m_{\tilde{t}}}{m_t}.
\eeq
This grows logarithmically with $m_{\tilde{t}}$.  For large
$m_{\tilde{t}}$ there is also a contribution to the Higgs
mass term that grows quadratically with $m_{\tilde{t}}$:
\beq
\De m_H^2 \sim \frac{N_c y_t^2}{16\pi^2} m_{\tilde t}^2.
\eeq
This large contribution must be tuned away to give the observed
value of the Higgs VEV, precisely the tuning
problem that SUSY is supposed to solve.
Satisfying the LEP Higgs mass bound necessitates a fine tuning of order
$1\%$, and the tuning increases exponentially with the Higgs mass.

There are a number of approaches to addressing this problem.
One is to extend the MSSM to get additional contributions to the quartic
coupling that are not fine tuned
\cite{Haber:1986gz,Drees:1987tp,Babu:1987kp,Batra:2003nj}.
Another approach is to extend the MSSM so that the Higgs decays
in a non-standard way, weakening the experimental limit $m_h > 114\GeV$
\cite{Dermisek:2005ar,Dermisek:2005gg,CFW,Carpenter:2007zz,Csaba1,Csaba2}.
Within the MSSM, \Ref{Kitano:2005wc}
argues that a specific region of parameter space
with large $A$ terms has small fine tuning,
while \Ref{GR} argues that anthropic considerations may explain
the fine tuning.
Some have advocated a large coefficient for the $S H_u H_d$ term of the NMSSM as
a remedy for the fine-tuning\cite{Barbieri:2006bg}.
Indeed, this can give an appreciable contribution to the Higgs boson mass but
at the expense of a Landau Pole at a low scale.
However, see \cite{Harnik:2003rs,Barbieri:2007tu, Birkedal:2004zx, Chang:2004db} for approaches to make this consistent with unification.
But the most popular approach by far is simply to ignore the problem
and study the fine-tuned MSSM.

In this paper we take this naturalness problem seriously
and further investigate non-standard Higgs decays
as a possible solution.
Searches for many non-standard Higgs decays have been performed, and
many are almost as sensitive as the search for Standard Model Higgs
(see \Ref{Higgsdecay} for a review).
We will focus on the cascade decay channel  $h \to aa \to 4j$,
which has significantly weaker limits than standard Higgs decays.
The strongest published limits on this decay come from the OPAL
experiment at LEP \cite{Abbiendi:2002in}.
The search was designed for $h \to jj$ and the jets from a light $a$ decay often mimic a single jet, so as a result these bounds exclude only light $a$ masses, $m_a \lesssim 10\GeV$ for $m_h < 86$ GeV.
There is also a model-independent limit $m_h > 82\GeV$ from OPAL
for $Zh$ production, looking for the $Z$ recoiling against an arbitrary
final state \cite{Abbiendi:2002qp}.
It is likely that a dedicated $h \to 4j$ search at LEP will give stronger
constraints.
However, no published result exists, and it is unclear whether a small
excess in this channel would have been noticed.

Is it natural for $h \to aa \to 4j$ to dominate?
Since $y_b^2 \sim 10^{-3}$, 
it is not difficult to construct models where another 2-body channel
such as $h\to aa$ dominates over $h\to \bar{b} b$.  
However, ensuring that $a \to jj$ is the dominant $a$ decay is more challenging.
Decays to quarks require flavor-violating couplings of the $a$,
and the decay to the heaviest quark generally dominates.
Such a scenario does not significantly reduce the experimentally
allowed Higgs mass since there are strong limits from LEP on
$h \to 4 b$ \cite{Schael:2006cr}.
We therefore focus on the possibility that $a \to gg$ dominates.
This decay can arise from the non-renormalizable operator
\begin{equation}
\label{eqn:aaNon}
\De{\mathcal L}  = \frac{1}{\La} a \tilde{G}_{\mu \nu} G^{\mu \nu},
\end{equation}
where we assume that $a$ is a pseudoscalar.
This coupling can be generated by a loop diagram involving
colored fields with a Yukawa coupling to $a$.
The result is that the partial width is suppressed by both a loop
factor and a heavy scale.
This suppression is potentially problematic
since the $a$ field can mix with the pseudoscalar Higgs boson of the MSSM,
and so the suppressed decay $a \to jj$ must compete with the mixing induced $a \to \bar{b}b$. 
Because of this, the simplest models will not give rise to $h \to 4g$.
For example, the NMSSM with the addition of the interaction \Eq{eqn:aaNon}
will not give the desired phenomenology.
In that model $m_a <  m_h/2$ only near the R-symmetric or Peccei-Quinn symmetric
limits.
In both cases the $a$ lives partially in the SM Higgs fields, and so
the tree level decay to quarks dominates over the loop suppressed decays
to gluons.

There are already SUSY models in the literature where $h \to 4 g$ dominates. 
The pioneering work is \Ref{CFW}, in which $a$ is the pseudoscalar
in a gauge singlet superfield $S$.
While this model is technically natural, allows $m_a$ up to  $m_h/2$,
and represents a proof-of-principle, it has a some undesirable
features.
For example, it has a UV divergent tadpole for $S$ and
requires non-standard soft SUSY breaking terms.
\Ref{Csaba1} constructed a SUSY little Higgs theory in which
$a$ is a pseudo Nambu-Goldstone boson (PNGB).
In this model $a \to gg$ dominates only if $a \to \bar{b}b$ is
kinematically forbidden.
This model is rather elaborate; 
it requires an extension of the Standard Model gauge group at the
TeV scale,
new flavor-dependent couplings,
and large Yukawa couplings with Landau poles not far above
the TeV scale.

In this paper we construct a simple and natural
model in which $h \to aa \to 4g$ dominates.
Our model extends the Higgs sector with gauge singlet superfields,
and the decay $a \to gg$ is mediated by additional vector-like
colored fields.
The $a$ is the PNGB
of an approximate $U(1)$ global symmetry, and can therefore be
naturally light.
The decay $a \to gg$ naturally dominates over $a \to \bar{b}b$
because the latter is automatically suppressed by additional
powers of explicit $U(1)$ breaking.
UV divergent tadpoles and mixing terms for the singlets are forbidden
by symmetries.
The model works for $\mbox{GeV} \lsim m_a \lsim \frac 12 m_h$,
motivating experimental searches over the full kinematically
allowed range.
Furthermore, our model has none of the undesirable features
of the previous models for $h \to 4g$.
It is compatible both with grand unification and standard
soft SUSY breaking terms.
The model also has no dimensionful SUSY invariant couplings,
and therefore preserves the solution of the $\mu$ problem
of the next-to-minimal supersymmetric standard model.

This paper also points out a possible ``smoking-gun'' signature of models
in which the decay $h \to aa \to 4g$ dominates.
The signal arises from the production of the colored particles
$X$ that mediate the $a \to gg$ decay.
The TeV scale is the natural scale of this theory, so the $X$ will be copiously produced at the LHC. 
The $X$ particles can decay only via flavor-dependent couplings.
These must be highly suppressed because of flavor bounds,
motivating (but not requiring) that they are stable on collider
scales.
If this is the case they will appear as $R$-hadrons at the LHC.
This is a possible early signal at the LHC, but it is certainly
not unique to our model
(see \cite{FairbairnRHadron} for a review and list of references).
The novel observation made here is that
if $R$-hadrons arise from mediators of $a \to gg$,
then there is a significant cross section for $a$
production in association with the $R$-hadrons,
{\it i.e.\/}\ $X \bar{X} a$ final states.
Unfortunately, $a \to gg$ is probably not observable at the LHC in such events. However, gauge coupling unification suggests that the new colored
fields are embedded in GUT multiplets, in which case they will
be electrically charged.
This gives a branching ratio for
$a \to \ga\ga$ of order $1\%$.
One can then search for $X \bar{X} a \rightarrow X \bar{X} \ga \ga$.
This may be observed with as little as 10 fb$^{-1}$ at the LHC.
This signal directly probes the $X \bar{X} a$ coupling,
and together with the non-observation of a standard model 
Higgs boson
would provide strong evidence that the $R$-hadrons arise from
colored particles that mediate exotic Higgs decays.

\section{A PNGB Model}
The model extends the MSSM with gauge singlet superfields
$S$, $N$, and $\bar{N}$,
with an approximate global $U(1)$ symmetry under
which $N$ and $\bar{N}$ have opposite charge.
In the $U(1)$ symmetry limit,
the $a$ to which the Higgs boson decays is contained in the $N$ and
$\bar{N}$ fields (see \Eq{eqn:Ns} below).
In addition, there is a vector-like pair of colored
superfields $X$ and $\bar{X}$.
The $U(1)$ invariant terms in the superpotential are
\beq
W = \la_H S H_u H_d
+ \frac{\ka_S}{3} S^3
+ \la_N S \bar{N} N
+ y_X N \bar{X} X.
\eeq
This is $U(1)$ invariant if $\bar{X}X$ carries $U(1)$ charge.
The global $U(1)$ symmetry is broken explicitly down to
$Z_3$ by naturally small superpotential terms
\beq
\De W = \frac{\ka_N}{3} N^3 
+ \frac{\ka_{\bar{N}}}{3} \bar{N}^3.
\eeq
There are actually two unbroken $Z_3$ symmetries.
The first is a subgroup of the global $U(1)$
that acts only on $N$ and $\bar{N}$,
and the second is one in which all fields are 
re-phased by $e^{2i\pi/3}$.
These symmetries are preserved by all standard soft SUSY
breaking terms (scalar and gaugino masses and $A$ terms).
They forbid UV divergent tadpoles for the singlet fields,
as well as UV divergent kinetic mixing among them.
This is important because 
these effects generally make $a \to \bar{b}b$ dominate 
over $a \to gg$.
If these $Z_3$ symmetries are exact, the theory has cosmologically
dangerous domain walls, but very small explicit breaking
is sufficient to eliminate this problem \cite{Abel:1996cr}.
The Higgs fields all have nonzero VEVs,
spontaneously breaking the approximate global $U(1)$ and
giving a mass to the colored fields $X$ and $\bar{X}$.
For $\ka_{N}, \ka_{\bar{N}} \ll 1$,
one of the pseudoscalar
fields is a light PNGB
which is the $a$ particle to which the Higgs decays.

In this model the
Higgs decay $h \to aa$ can easily dominate over
$h \to \bar{b}b$.
We first give a discussion of this point in the
limit where explicit breaking of the global $U(1)$ vanishes.
It is convenient to parametrize the neutral Higgs fields by
\beq\label{Hudfields}
H_u = \frac{1}{\sqrt 2} \pmatrix{0 \cr
(v + h_v) s_\be + (H_v + i A_v) c_\be \cr},
\quad
H_d = \frac{1}{\sqrt 2} \pmatrix{
(v + h_v) c_\be - (H_v - i A_v) s_\be \cr 0 \cr},
\eeq
where $s_\be = \sin\be$ {\it etc.\/}
(see {\it e.g\/.}\ \cite{Dobrescu:2000yn}).
Here $h_v$ is not a mass eigenstate, but it is the
state that unitarizes $WW$ scattering at
high energies.
In the absence of fine tuning we expect one of the mass
eigenstates $h$ to be mostly $h_v$,
and we follow common practice by calling this ``the'' Higgs boson.
The excitations of $N$ and $\bar{N}$ are conveniently parametrized by
\beq
\label{eqn:Ns}
N = \frac{1}{\sqrt 2}(v_N + n) e^{i(A_n + a_n)/f_N},
\quad
\bar{N} = \frac{1}{\sqrt 2}(v_{\bar{N}} + \bar{n}) e^{i(A_n - a_n)/f_n},
\eeq
with $f_n = \sqrt{v_N^2 + v_{\bar{N}}^2}$.
In this parametrization the $a_n$ is derivatively
coupled in the $U(1)$ symmetry limit.
It also does not mix with any other field in this limit,
and can therefore be identified with the massless NGB
to which the Higgs decays.

The relevant coupling for $h \to aa$ decay comes from the kinetic terms
for $N$ and $\bar{N}$:
\beq
\label{eqn:RelevantCoupling}
\De\scr{L} = \frac{1}{f_n^2} (v_N n + v_{\bar{N}} \bar{n})
\partial^\mu a_n \partial_\mu a_n + \cdots.
\eeq
When the $n$ and $\bar{n}$ scalars mix with $h_{v}$, this term can mediate the desired Higgs boson decays.
The relevant decay width is given by:
\begin{equation}
\Gamma(h \to aa) =
\frac{m_{h}^{3}(v_{N}  U_{n h} + v_{\bar{N}} U_{\bar{n} h})^{2}}
{32 \pi f_{n}^{4}}
\left(1- \frac{2 m_a^{2}}{m_{h^{2}}}\right)^2 
\left(  1 - \frac{4 m_{a}^2}{m_{h}^{2}}\right)^{1/2},
\end{equation}
where $U_{nh}$ and $U_{\bar{n} h}$ are the mixings between the Higgs boson
(the mass eigenstate with the largest overlap with $h_{v}$)
and the $n$ and $\bar{n}$ fields.

To get a concrete estimate for the branching ratio $h \to aa$,
it is useful to define the combination
\beq
n_{+} = \frac{1}{\sqrt{2}}( n + \bar{n}).
\eeq
In the limit that $v_{N} \simeq v_{\bar{N}}$ it is this combination that appears in the relevant coupling in Eq.~(\ref{eqn:RelevantCoupling}).
It is this same combination that mixings with the Higgs boson via the
$|F_{S}|^2$ term in the scalar potential:
\beq
V_F = \frac{1}{\sqrt{2}} \la_H \la_N v \sin(2\be) v_N n_{+} h
+ \cdots.
\eeq
The mixing angle between the $n_{+}$ and the Higgs can then be estimated by dividing this result by the largest diagonal entry in the $2 \times 2$
mass matrix for $n_+$ and $h$,
which we assume to be $\lambda_N v_{N}$.
In the limit $v_N \sim v_{\bar{N}} \sim f_n$ we then obtain
\beq
\Ga(h \to aa) \sim \frac{1}{16\pi} \frac{m_h^3}{v_N^2} \left( \frac{\la_H v \sin(2\be)}{2 \la_N v_N} \right)^2.
\eeq
For order-1 values of the couplings, $\tan{\be}$ not too large,
and $f_n \sim v$ this can easily dominate over $h \to \bar{b}b$.

We now discuss the $a$ decays.
In the limit where $a$ is an exact NGB, the only coupling
linear in $a$ is a coupling to fields that are charged under
the global $U(1)$, {\it i.e.\/}\ $X$ and $\bar{X}$.
This means that the $agg$ coupling is unsuppressed in the
$U(1)$ symmetry limit.
Decays such as $a \to \bar{b}b$
occur at tree level due to mixing of $a$ and the Higgs pseudoscalar
$A_v$, but this mixing is suppressed
by the small explicit $U(1)$ breaking couplings $\ka_{N}$, $\ka_{\bar{N}}$.
This is the basic reason that $a \to gg$ can naturally dominate
even though it is loop suppressed.
More precisely, assuming $\ka_N, \ka_{\bar{N}} \sim \ep \ll 1$,
we have 
\beq
\Ga(a \to gg) = \frac{9 h_{X\bar{X}a}^2 b_X^2(N_c^2-1)}{64\pi} \left( \frac{g_3^2}{16\pi^2} \right)^2 \frac{m_a^3}{(y_N v_N/\sqrt{2})^2} .
\eeq
On the other hand
\beq
\Ga(a \to \bar{b}b) \sim \frac{N_c y_b^2}{16\pi} \ep^2 m_a.
\eeq
Assuming that all masses and VEVs are of the same order,
we have $m_a^2 \sim \ep v^2$,
and therefore
\beq
\frac{\Ga(a \to \bar{b}b)}{\Ga(a \to gg)} 
\sim \frac{\ep y_b^2}{N_c} \left( \frac{g_3^2}{16\pi^2} \right)^{-2}.
\eeq
Therefore $a \to \bar{b}b$
is sub-dominant for sufficiently small $\ep$.
Performing the full calculation, we find choices of parameters where
$\ep$ can be large enough that $m_{a} \simeq 40\GeV$,
while $a \to gg$ still dominates.

We have not undertaken a exhaustive parameter scan of this model,
but it is not hard to find phenomenologically acceptable
benchmark points with no fine tuning.
We present one such point for concreteness.
The relevant values are shown in Table \ref{table:benchmark}.
We have used the VEVs as input parameters, with the soft
mass parameters as output parameters.
\begin{table}
\begin{tabular} {lllll}
\hline
$\tan \beta =6$ & $\lambda_H = 0.13$ & $\kappa_{S} =-0.42$ & $\lambda_{N}=0.38$  & $\kappa_{N} = 0.001$ \\
\hline
$A_{\kappa_N} = 0 \rm{\; GeV}$ & $A_{\lambda_H} = -75 \rm{\; GeV}$ & $A_{\kappa_S} = 200 \rm{\;GeV}$&
$A_{\lambda_N} = -65  \rm{\; GeV}$\\
\hline
$v_{s} = -1220 \rm{\; GeV}$ & $v_{N} = -425 \rm{\; GeV}$ & $v_{\bar{N}}= -170 \rm{\; GeV}$ \\
\hline
$y_{X} = 1.0$ & $\tilde{m}_{X}^2 = (250$ GeV$)^2$ \\
\hline
\end{tabular}
\caption{A set of benchmark values resulting in $h \to 4g$ decays.}
\label{table:benchmark}
\end{table}

This point results in a mass spectrum and branching ratios as given in Table \ref{table:pheno}.  LEP searches constrain
$\xi_h^2 BR(h \to \bar{b} b) \lesssim 0.05$
for $m_h < 90$ GeV,
and $\xi_h^2 BR(h \to \bar{b} b) \lesssim 0.1- 0.2$ for 90 GeV $<m_h<$ 100 GeV  \cite{Schael:2006cr}, where
\beq
\xi_h = \frac{g_{hZZ}}{g_{hZZ}^{({\rm SM})}}\,.
\eeq
Therefore only a very modest contribution from the stop squarks is
required to give a Higgs mass above the LEP search limits for $h \to \bar{b} b$.

\begin{table}
\begin{tabular} {lllll}
\hline
$m_{h_1} = 78 \rm{\; GeV}$ & $m_{h_2} = 88 \rm{\; GeV}$ & $m_{h_3} = 538 \rm{\; GeV}$ & $m_{h_4} = 593 \rm{\; GeV}$ & $m_{h_5} = 709 \rm{\; GeV}$ \\
\hline
$m_{a_1} = 17 \rm{\; GeV}$ & $m_{a_2} = 455 \rm{\; GeV}$ & $m_{a_3} =541 \rm{\; GeV}$ & $m_{a_4} = 663 \rm{\; GeV}$ \\
\hline
$m_{X} =  300 \rm{\; GeV}$  \\
\hline
$\xi^2_{h_1} = 0.18$ & $\xi^2_{h_1} BR(h_1 \to \bar{b}b) = 0.0035$ & $\xi^2_{h_1} BR(h_1 \to aa) = 0.18$ & $BR(a \to gg) = 1$ \\
\hline
$\xi^2_{h_2} = 0.81$ & $\xi^2_{h_2} BR(h_2 \to \bar{b}b) = 0.14$ & $\xi^2_{h_2} BR(h_2 \to aa) = 0.66$ \\
\hline
\end{tabular}
\caption{Tree Level mass spectrum for the benchmark of Table \ref{table:benchmark}.}
\label{table:pheno}
\end{table}

\section {Phenomenology}
We now turn to the phenomenology of the model.
After a brief discussion of superpartner and Higgs searches,
we turn to signals involving the mediator fields $X$ and $\bar{X}$,
where we identify a new possible distinctive signature of this class
of models.

The motivation for this model is that fine-tuned radiative corrections
from a large stop mass are not required.
Therefore, we expect all superpartners to be near their current
experimental bounds, so standard SUSY searches are expected to find
superpartners early at the LHC.

Discovering the Higgs is of course
more difficult in models where $h \to 4g$ dominates.
There are recent analyses that claim that
$h \to 4 g$ may be observable using jet substructure methods
with 100~fb$^{-1}$ of LHC luminosity at 14 TeV \cite{Falkowski:2010hi,Chen:2010wk}.
These techniques were studied for $m_a < 2 m_b$,
but in the model we are discussing $m_a$ can be as large as $m_h/2$
and these search strategies are not expected to be effective in this case.
For $m_a$ near $m_h/2$, one may be able to use
a boosted Higgs as studied in \cite{Falkowski:2010hi},
since the four gluon jets may appear as a single fat jet.
The model described in this paper motivates a detailed experimental study of the
$h \to aa \to 4g$ final state over the entire allowed kinematic range. 

\begin{figure}
\begin{center}
\scalebox{0.75}{\includegraphics{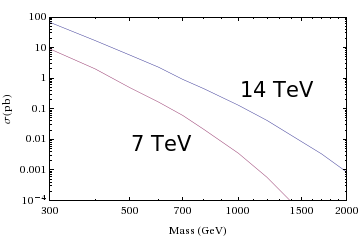}}
\end{center}
\caption{Cross section for $X$ fermion production at the LHC for different center of mass energies.
\label{fig:X-hadronproduction}}
\end{figure}

We now turn to the phenomenology of the $X$ mediators.
They must be below the TeV scale because they get their mass from the 
VEV of $N$, and since they are colored they will be copiously produced
at the LHC.
For definiteness we will assume that the fermion components of
$X$ are lighter than the scalar components
({\it i.e.\/}\ the soft mass-squared terms for the $X$ scalars are
positive),
but our main points do not depend on this assumption.
The production cross section for a color triplet fermions
is shown in figure \ref{fig:X-hadronproduction}.
Color conservation implies that $X$ fermions must decay into 
an odd number of quarks, so the
decay necessarily violates flavor.
For example, if the mediators are part of a $\bf{5} \oplus \bar{\bf 5}$
of $SU(5)$ (to preserve gauge coupling unification)
$X$ will have the quantum numbers of a right-handed down quark.
In this case the masses of these fermions come from the superpotential terms
\beq\label{epsYuk}
\De W = y_X N \bar{X} X  + \ep_i Q_i H_d X,
\eeq
where $i = 1, 2, 3$ is a flavor index.
We have rotated away a possible
$N \bar{X} d^c_i$ term by a field redefinition of the
fields $d_i^c$ and $X$.  
The Yukawa couplings $\ep_i$ mix the $X$ with the $d_i^c$,
allowing weak decays of the heavy mediators.
Note that the dominant mass term $\bar{X}X$ preserves electroweak
symmetry, and therefore does not give rise to large corrections
to electroweak precision observables.
The couplings $\ep_i$ can be made small enough to satisfy constraints
from flavor and precision electroweak observables while giving
prompt decays.
This is a well-motivated scenario that gives rise
to ``fourth generation'' phenomenology without unnatural
tuning to satisfy experimental constraints.

However, it is also possible that Yukawa couplings such as
the $\ep_i$ in
\Eq{epsYuk} are not present or are highly suppressed.
For example, they are forbidden if $X$
is even under $R$-parity.
In this case, the leading interaction that can give rise to
$X$ decays are higher-dimension operators such as
$\De W \sim (\bar{X} d^c)(L H_u)$.
This can easily give an $X$ that is stable on collider scales:
for example if the scale suppressing such higher-dimension operators is
the GUT scale, $X$ has a decay length of order $10^6$~km.

Stable $X$ particles will hadronize and appear in the detector as
``$X$-hadrons'' similar to $R$-hadrons arising from SUSY models with
gluino or squark LSP \cite{Raby:1997pb,Baer:1998pg,Mafi:2000kg,Raby:1997ba,Mafi:1999dg}
or split supersymmetry
\cite{ArkaniHamed:2004fb}.
These ``X-hadrons" may appear as highly ionizing charged tracks.
The acceptance for  $R$-hadrons at CMS is expected to be $\sim 25\%$
\cite{CMSRHadrons}, so they can be observed in early LHC running.
Early searches at CMS based on only 198 nb$^{-1}$ of data for such exotics
already place a weak bound on this scenario.
An interpolation between the results of \Ref{CMSRHadrons} indicates a bound
somewhat less than 200 GeV.
This bound should improve substantially soon.
Another possibility is that
the X hadrons may stop in the detector and decay much later \cite{latedecay},
Existing searches for this signal are also sensitive to this model
\cite{Abazov:2007ht,Collaboration:2010uf,Farrar:2010ps}.
Present searches for stopped gluinos can be reinterpreted as a bound on stopping X-hadrons, with a bound of somewhat less than 300 GeV, depending on what assumptions are made about the stopping probability of the X-hadrons
\cite{Collaboration:2010uf}.

If $R$-hadrons or ``fourth generation'' quarks
are discovered at the LHC, the next question will be
what model they come from.
Here we point out that a direct confirmation of the present model
can come from $a$ radiation from an $X$ particle,
since this directly probes the $a \bar{X}X$ coupling responsible
for the second stage of the Higgs decay. This is given by
\beq
g_{a\bar{X}X} = \frac{y_X v_N}{\sqrt{2} f_n}.
\eeq
The point is that discovery of either signal gives a 
sample of events that are essentially free of standard model backgrounds.
The $a$ decays mostly to low $p_T$ gluon jets, and so a signature of $X\bar{X}a$
production could be $X\bar{X}jj$.
However, resolving the $a$ peak in the $jj$ invariant mass distribution 
appears to be impossible because the jets have $p_T \lsim 50$~GeV,
and the energy resolution is very poor for such jets.
There is a more promising signal if the $X$ also carries electric
charge, for example if $X$ is a {\bf 5} under $SU(5)$.
Then $a \to \ga\ga$ is also allowed, and we can search for
$X\bar{X}\ga\ga$.
If all scalars and fermions have a common mass, then
\beq \label{eq:br2gamma}
\mbox{BR}(a \to \gamma \gamma) =3.7 \times 10^{-3}.
\eeq 
Backgrounds from radiation of photons or jet faking photons are negligible
(as are all standard model backgrounds).  

In Table \ref{tab:XXa} we show the production cross section for different masses
of the $X$ and $a$ at the LHC with 14 TeV.
We assume that the signal acceptance is close to that of the $X$-hadrons
($\simeq 25\%$)
and use the branching ratio of \Eq{eq:br2gamma}.
We see that we can get a handful of events in as little as 10 fb$^{-1}$,
and a large part of the parameter space can be discovered
in 100 fb$^{-1}$.

\begin{table}
\begin{tabular}{|c|c|c|c|}
\hline
$m_X$ (GeV) & $m_a$ (GeV)  & $\sigma(pp \to X\bar{X}a)$  & 
$\mbox{eff} \times \sigma \times \mbox{BR}(a \to \gamma \gamma)$ \\
\hline
300         & 15           &  3100      & 2.9 \\
300         & 30           &  1800     & 1.7\\
400         & 15           &  870     & 0.80\\
400         & 30           &  510     & 0.47\\
500         & 15           &  300     & 0.28\\
500         & 30           &  170     & 0.16\\
700         & 15           &  47    & 0.043\\
700         & 30           &  29    & 0.027\\
\hline
\end{tabular}
\caption{$\sigma(pp\to X\bar{X}a)$ in fb
for different masses of the $X$ and $a$,
assuming the Yukawa coupling of $X$ to $a$ is 1,
calculated using MadGraph \cite{Madgraph}.
The last column gives the expected cross section for the signal
$X\bar{X}\gamma\gamma$ multiplied by a signal efficiency of 0.25 \cite{CMSRHadrons}.   \label{tab:XXa}}
\end{table}

The heavier superpartner of the $X$ particle (assumed here to be a scalar)
also has interesting phenomenology.
It decays to the lighter $X$ particle by emitting a gluino (possibly virtual)
or neutralino.
If the gluino is lighter than the $X$ partner this will give rise to a
pair of $R$-hadrons together with a full SUSY cascade initiated by the gluinos.
If this decay is kinematically forbidden then the decay to the LSP is expected to
dominate, and we get a pair of $R$-hadrons plus missing energy.
These striking events would directly show that the $X$-hadrons have a new
$R$-parity odd superpartner.
(This contrasts with $R$-hadrons from gluinos of squarks, where the superpartner
is an ordinary particle.)
In the case where the $X$ scalar is lighter than squarks and gluinos,
such processes could even be the dominant source of missing energy.
Thus, dedicated searches for $R$-hadrons in association with missing
energy may be called for if searches for either $R$-hadrons or SUSY
see signs of a signal. 

Although their existence is not necessary for hiding the Higgs boson, unification suggests the existence of electroweak doublet partners of in the $X$ multiplet.
These might also lead to interesting collider signatures
\cite{ThomasWells, RandallBuckley}, but the electrically charged ``lepton''
in the doublet is expected to rapidly decay to a soft pion and the
electrically neutral ``heavy neutrino'', leading to a challenging signature.
With sufficient luminosity, one might observe associated production
of $a$ with these ``leptons'' in the final state
$\ga\ga$ plus missing $E_T$, giving additional evidence for this mechanism.

\section{Conclusions}
We have constructed a simple supersymmetric
model in which the Higgs naturally decays dominantly via
$h \to aa \to 4g$.
This allows $m_h < m_Z$ and completely eliminates the need for
fine tuning to satisfy the LEP Higgs bounds.
Models of this kind have been considered previously in the
literature, and the main difficulty is getting $a \to gg$
to dominate over $a \to \bar{f}f$ where $f$ is the heaviest
kinematically accessible fermion (generally $b$ or $\tau$).
The main ingredient in the present model
is that $a$ is a pseudo Nambu-Goldstone
boson associated with a spontaneously broken $U(1)$ symmetry
in the Higgs sector.
The decay $a \to \bar{f}f$ is suppressed by
small explicit breaking of the $U(1)$ symmetry, while
the coupling to gluons is not.
Our model is significantly simpler than existing models in the literature,
and it works for $a$ masses in the full kinematically
allowed range.
We believe this model provides strong additional motivation
for searching for the Higgs in this channel.
This is especially important for larger $a$ masses where
current search strategies become ineffective.

We also pointed out a new ``smoking gun'' signature
for models in which $h \to 4g$ is the dominant Higgs
decay mode.
These models necessarily have colored particles $X$ with a
large coupling to $a$ to mediate $a \to gg$.
The masses of the $X$ particles must be below the TeV scale
in order for $a \to gg$ to be large enough, so $X$ particles
can be copiously produced at LHC.
They can decay only through flavor-violating couplings,
and therefore may be stable on collider scales.
In this case, they appear in the detector as 
heavy stable colored particles, ``$X$-hadrons.''
Alternatively, if they decay, the most natural possibility
is weak decays similar to a fourth generation.
In either case, we expect discovery of $X$ particles with
a large number of events.
This in turn gives a nearly background-free sample of $X$
production events, and we can look for the associated production
of $a$ with $X$ pairs.
This directly probes the coupling responsible for $a \to gg$.
Associated production with $a \to gg$ is very difficult to
observe, but $a \to \ga\ga$ is expected to have a $\sim 1\%$
branching ratio, and is readily observable.
Additionally, observation of $X$-hadrons with missing energy
is a direct sign that the $X$ particle has a new
$R$-parity odd superpartner.

\acknowledgments{We acknowledge helpful conversations with
S. Chang, C. Csaki, D. Poland,  T. Volansky, I. Yavin, and M. Chertok.
We thank the Aspen Center for Physics, where this work was initiated.
ML and DP are supported by the Department of Energy
under grant DE-FG02-91-ER40674.
AP is supported in part by NSF CAREER Grant NSF-PHY-0743315,
and in part by the Department of Energy under grant DE-FG02-95-ER40899.}

\bibliography{hiddenhiggs}

\begin{thebibliography}{41}
\expandafter\ifx\csname natexlab\endcsname\relax\def\natexlab#1{#1}\fi
\expandafter\ifx\csname bibnamefont\endcsname\relax
  \def\bibnamefont#1{#1}\fi
\expandafter\ifx\csname bibfnamefont\endcsname\relax
  \def\bibfnamefont#1{#1}\fi
\expandafter\ifx\csname citenamefont\endcsname\relax
  \def\citenamefont#1{#1}\fi
\expandafter\ifx\csname url\endcsname\relax
  \def\url#1{\texttt{#1}}\fi
\expandafter\ifx\csname urlprefix\endcsname\relax\def\urlprefix{URL }\fi
\providecommand{\bibinfo}[2]{#2}
\providecommand{\eprint}[2][]{\url{#2}}

\bibitem[{\citenamefont{Dimopoulos et~al.}(1981)\citenamefont{Dimopoulos, Raby,
  and Wilczek}}]{DRW}
\bibinfo{author}{\bibfnamefont{S.}~\bibnamefont{Dimopoulos}},
  \bibinfo{author}{\bibfnamefont{S.}~\bibnamefont{Raby}}, \bibnamefont{and}
  \bibinfo{author}{\bibfnamefont{F.}~\bibnamefont{Wilczek}},
  \bibinfo{journal}{Phys. Rev.} \textbf{\bibinfo{volume}{D24}},
  \bibinfo{pages}{1681} (\bibinfo{year}{1981}).

\bibitem[{\citenamefont{Haber and Sher}(1987)}]{Haber:1986gz}
\bibinfo{author}{\bibfnamefont{H.~E.} \bibnamefont{Haber}} \bibnamefont{and}
  \bibinfo{author}{\bibfnamefont{M.}~\bibnamefont{Sher}},
  \bibinfo{journal}{Phys.Rev.} \textbf{\bibinfo{volume}{D35}},
  \bibinfo{pages}{2206} (\bibinfo{year}{1987}).

\bibitem[{\citenamefont{Drees}(1987)}]{Drees:1987tp}
\bibinfo{author}{\bibfnamefont{M.}~\bibnamefont{Drees}},
  \bibinfo{journal}{Phys.Rev.} \textbf{\bibinfo{volume}{D35}},
  \bibinfo{pages}{2910} (\bibinfo{year}{1987}).

\bibitem[{\citenamefont{Babu et~al.}(1987)\citenamefont{Babu, He, and
  Ma}}]{Babu:1987kp}
\bibinfo{author}{\bibfnamefont{K.}~\bibnamefont{Babu}},
  \bibinfo{author}{\bibfnamefont{X.-G.} \bibnamefont{He}}, \bibnamefont{and}
  \bibinfo{author}{\bibfnamefont{E.}~\bibnamefont{Ma}},
  \bibinfo{journal}{Phys.Rev.} \textbf{\bibinfo{volume}{D36}},
  \bibinfo{pages}{878} (\bibinfo{year}{1987}).

\bibitem[{\citenamefont{Batra et~al.}(2004)\citenamefont{Batra, Delgado,
  Kaplan, and Tait}}]{Batra:2003nj}
\bibinfo{author}{\bibfnamefont{P.}~\bibnamefont{Batra}},
  \bibinfo{author}{\bibfnamefont{A.}~\bibnamefont{Delgado}},
  \bibinfo{author}{\bibfnamefont{D.~E.} \bibnamefont{Kaplan}},
  \bibnamefont{and} \bibinfo{author}{\bibfnamefont{T.~M.} \bibnamefont{Tait}},
  \bibinfo{journal}{JHEP} \textbf{\bibinfo{volume}{0402}}, \bibinfo{pages}{043}
  (\bibinfo{year}{2004}), \eprint{hep-ph/0309149}.

\bibitem[{\citenamefont{Dermisek and Gunion}(2005)}]{Dermisek:2005ar}
\bibinfo{author}{\bibfnamefont{R.}~\bibnamefont{Dermisek}} \bibnamefont{and}
  \bibinfo{author}{\bibfnamefont{J.~F.} \bibnamefont{Gunion}},
  \bibinfo{journal}{Phys. Rev. Lett.} \textbf{\bibinfo{volume}{95}},
  \bibinfo{pages}{041801} (\bibinfo{year}{2005}), \eprint{hep-ph/0502105}.

\bibitem[{\citenamefont{Dermisek and Gunion}(2006)}]{Dermisek:2005gg}
\bibinfo{author}{\bibfnamefont{R.}~\bibnamefont{Dermisek}} \bibnamefont{and}
  \bibinfo{author}{\bibfnamefont{J.~F.} \bibnamefont{Gunion}},
  \bibinfo{journal}{Phys. Rev.} \textbf{\bibinfo{volume}{D73}},
  \bibinfo{pages}{111701} (\bibinfo{year}{2006}), \eprint{hep-ph/0510322}.

\bibitem[{\citenamefont{Chang et~al.}(2006)\citenamefont{Chang, Fox, and
  Weiner}}]{CFW}
\bibinfo{author}{\bibfnamefont{S.}~\bibnamefont{Chang}},
  \bibinfo{author}{\bibfnamefont{P.~J.} \bibnamefont{Fox}}, \bibnamefont{and}
  \bibinfo{author}{\bibfnamefont{N.}~\bibnamefont{Weiner}},
  \bibinfo{journal}{JHEP} \textbf{\bibinfo{volume}{08}}, \bibinfo{pages}{068}
  (\bibinfo{year}{2006}), \eprint{hep-ph/0511250}.

\bibitem[{\citenamefont{Carpenter et~al.}(2007)\citenamefont{Carpenter, Kaplan,
  and Rhee}}]{Carpenter:2007zz}
\bibinfo{author}{\bibfnamefont{L.~M.} \bibnamefont{Carpenter}},
  \bibinfo{author}{\bibfnamefont{D.~E.} \bibnamefont{Kaplan}},
  \bibnamefont{and} \bibinfo{author}{\bibfnamefont{E.-J.} \bibnamefont{Rhee}},
  \bibinfo{journal}{Phys. Rev. Lett.} \textbf{\bibinfo{volume}{99}},
  \bibinfo{pages}{211801} (\bibinfo{year}{2007}), \eprint{hep-ph/0607204}.

\bibitem[{\citenamefont{Bellazzini et~al.}(2009)\citenamefont{Bellazzini,
  Csaki, Falkowski, and Weiler}}]{Csaba1}
\bibinfo{author}{\bibfnamefont{B.}~\bibnamefont{Bellazzini}},
  \bibinfo{author}{\bibfnamefont{C.}~\bibnamefont{Csaki}},
  \bibinfo{author}{\bibfnamefont{A.}~\bibnamefont{Falkowski}},
  \bibnamefont{and} \bibinfo{author}{\bibfnamefont{A.}~\bibnamefont{Weiler}},
  \bibinfo{journal}{Phys.Rev.} \textbf{\bibinfo{volume}{D80}},
  \bibinfo{pages}{075008} (\bibinfo{year}{2009}), \eprint{arXiv:0906.3026}.

\bibitem[{\citenamefont{Bellazzini et~al.}(2010)\citenamefont{Bellazzini,
  Csaki, Falkowski, and Weiler}}]{Csaba2}
\bibinfo{author}{\bibfnamefont{B.}~\bibnamefont{Bellazzini}},
  \bibinfo{author}{\bibfnamefont{C.}~\bibnamefont{Csaki}},
  \bibinfo{author}{\bibfnamefont{A.}~\bibnamefont{Falkowski}},
  \bibnamefont{and} \bibinfo{author}{\bibfnamefont{A.}~\bibnamefont{Weiler}},
  \bibinfo{journal}{Phys.Rev.} \textbf{\bibinfo{volume}{D81}},
  \bibinfo{pages}{075017} (\bibinfo{year}{2010}), \eprint{arXiv:0910.3210}.

\bibitem[{\citenamefont{Kitano and Nomura}(2005)}]{Kitano:2005wc}
\bibinfo{author}{\bibfnamefont{R.}~\bibnamefont{Kitano}} \bibnamefont{and}
  \bibinfo{author}{\bibfnamefont{Y.}~\bibnamefont{Nomura}},
  \bibinfo{journal}{Phys.Lett.} \textbf{\bibinfo{volume}{B631}},
  \bibinfo{pages}{58} (\bibinfo{year}{2005}), \eprint{hep-ph/0509039}.

\bibitem[{\citenamefont{Giudice and Rattazzi}(2006)}]{GR}
\bibinfo{author}{\bibfnamefont{G.~F.} \bibnamefont{Giudice}} \bibnamefont{and}
  \bibinfo{author}{\bibfnamefont{R.}~\bibnamefont{Rattazzi}},
  \bibinfo{journal}{Nucl. Phys.} \textbf{\bibinfo{volume}{B757}},
  \bibinfo{pages}{19} (\bibinfo{year}{2006}), \eprint{hep-ph/0606105}.

\bibitem[{\citenamefont{Barbieri et~al.}(2007)\citenamefont{Barbieri, Hall,
  Nomura, and Rychkov}}]{Barbieri:2006bg}
\bibinfo{author}{\bibfnamefont{R.}~\bibnamefont{Barbieri}},
  \bibinfo{author}{\bibfnamefont{L.~J.} \bibnamefont{Hall}},
  \bibinfo{author}{\bibfnamefont{Y.}~\bibnamefont{Nomura}}, \bibnamefont{and}
  \bibinfo{author}{\bibfnamefont{V.~S.} \bibnamefont{Rychkov}},
  \bibinfo{journal}{Phys. Rev.} \textbf{\bibinfo{volume}{D75}},
  \bibinfo{pages}{035007} (\bibinfo{year}{2007}), \eprint{hep-ph/0607332}.

\bibitem[{\citenamefont{Harnik et~al.}(2004)\citenamefont{Harnik, Kribs,
  Larson, and Murayama}}]{Harnik:2003rs}
\bibinfo{author}{\bibfnamefont{R.}~\bibnamefont{Harnik}},
  \bibinfo{author}{\bibfnamefont{G.~D.} \bibnamefont{Kribs}},
  \bibinfo{author}{\bibfnamefont{D.~T.} \bibnamefont{Larson}},
  \bibnamefont{and} \bibinfo{author}{\bibfnamefont{H.}~\bibnamefont{Murayama}},
  \bibinfo{journal}{Phys. Rev.} \textbf{\bibinfo{volume}{D70}},
  \bibinfo{pages}{015002} (\bibinfo{year}{2004}), \eprint{hep-ph/0311349}.

\bibitem[{\citenamefont{Barbieri et~al.}(2008)\citenamefont{Barbieri, Hall,
  Papaioannou, Pappadopulo, and Rychkov}}]{Barbieri:2007tu}
\bibinfo{author}{\bibfnamefont{R.}~\bibnamefont{Barbieri}},
  \bibinfo{author}{\bibfnamefont{L.~J.} \bibnamefont{Hall}},
  \bibinfo{author}{\bibfnamefont{A.~Y.} \bibnamefont{Papaioannou}},
  \bibinfo{author}{\bibfnamefont{D.}~\bibnamefont{Pappadopulo}},
  \bibnamefont{and} \bibinfo{author}{\bibfnamefont{V.~S.}
  \bibnamefont{Rychkov}}, \bibinfo{journal}{JHEP}
  \textbf{\bibinfo{volume}{03}}, \bibinfo{pages}{005} (\bibinfo{year}{2008}),
  \eprint{0712.2903}.

\bibitem[{\citenamefont{Birkedal et~al.}(2005)\citenamefont{Birkedal, Chacko,
  and Nomura}}]{Birkedal:2004zx}
\bibinfo{author}{\bibfnamefont{A.}~\bibnamefont{Birkedal}},
  \bibinfo{author}{\bibfnamefont{Z.}~\bibnamefont{Chacko}}, \bibnamefont{and}
  \bibinfo{author}{\bibfnamefont{Y.}~\bibnamefont{Nomura}},
  \bibinfo{journal}{Phys.Rev.} \textbf{\bibinfo{volume}{D71}},
  \bibinfo{pages}{015006} (\bibinfo{year}{2005}), \eprint{hep-ph/0408329}.

\bibitem[{\citenamefont{Chang et~al.}(2005)\citenamefont{Chang, Kilic, and
  Mahbubani}}]{Chang:2004db}
\bibinfo{author}{\bibfnamefont{S.}~\bibnamefont{Chang}},
  \bibinfo{author}{\bibfnamefont{C.}~\bibnamefont{Kilic}}, \bibnamefont{and}
  \bibinfo{author}{\bibfnamefont{R.}~\bibnamefont{Mahbubani}},
  \bibinfo{journal}{Phys. Rev.} \textbf{\bibinfo{volume}{D71}},
  \bibinfo{pages}{015003} (\bibinfo{year}{2005}), \eprint{hep-ph/0405267}.

\bibitem[{\citenamefont{Chang et~al.}(2008)\citenamefont{Chang, Dermisek,
  Gunion, and Weiner}}]{Higgsdecay}
\bibinfo{author}{\bibfnamefont{S.}~\bibnamefont{Chang}},
  \bibinfo{author}{\bibfnamefont{R.}~\bibnamefont{Dermisek}},
  \bibinfo{author}{\bibfnamefont{J.~F.} \bibnamefont{Gunion}},
  \bibnamefont{and} \bibinfo{author}{\bibfnamefont{N.}~\bibnamefont{Weiner}},
  \bibinfo{journal}{Ann. Rev. Nucl. Part. Sci.} \textbf{\bibinfo{volume}{58}},
  \bibinfo{pages}{75} (\bibinfo{year}{2008}), \eprint{0801.4554}.

\bibitem[{\citenamefont{Abbiendi et~al.}(2003{\natexlab{a}})}]{Abbiendi:2002in}
\bibinfo{author}{\bibfnamefont{G.}~\bibnamefont{Abbiendi}} \bibnamefont{et~al.}
  (\bibinfo{collaboration}{OPAL Collaboration}), \bibinfo{journal}{Eur.Phys.J.}
  \textbf{\bibinfo{volume}{C27}}, \bibinfo{pages}{483}
  (\bibinfo{year}{2003}{\natexlab{a}}), \eprint{hep-ex/0209068}.

\bibitem[{\citenamefont{Abbiendi et~al.}(2003{\natexlab{b}})}]{Abbiendi:2002qp}
\bibinfo{author}{\bibfnamefont{G.}~\bibnamefont{Abbiendi}} \bibnamefont{et~al.}
  (\bibinfo{collaboration}{OPAL Collaboration}), \bibinfo{journal}{Eur.Phys.J.}
  \textbf{\bibinfo{volume}{C27}}, \bibinfo{pages}{311}
  (\bibinfo{year}{2003}{\natexlab{b}}), \eprint{hep-ex/0206022}.

\bibitem[{\citenamefont{Schael et~al.}(2006)}]{Schael:2006cr}
\bibinfo{author}{\bibfnamefont{S.}~\bibnamefont{Schael}} \bibnamefont{et~al.}
  (\bibinfo{collaboration}{ALEPH Collaboration, DELPHI Collaboration, L3
  Collaboration, OPAL Collaborations, LEP Working Group for Higgs Boson
  Searches}), \bibinfo{journal}{Eur.Phys.J.} \textbf{\bibinfo{volume}{C47}},
  \bibinfo{pages}{547} (\bibinfo{year}{2006}), \eprint{hep-ex/0602042}.

\bibitem[{\citenamefont{Fairbairn et~al.}(2007)\citenamefont{Fairbairn, Kraan,
  Milstead, Sjostrand, Skands et~al.}}]{FairbairnRHadron}
\bibinfo{author}{\bibfnamefont{M.}~\bibnamefont{Fairbairn}},
  \bibinfo{author}{\bibfnamefont{A.}~\bibnamefont{Kraan}},
  \bibinfo{author}{\bibfnamefont{D.}~\bibnamefont{Milstead}},
  \bibinfo{author}{\bibfnamefont{T.}~\bibnamefont{Sjostrand}},
  \bibinfo{author}{\bibfnamefont{P.~Z.} \bibnamefont{Skands}},
  \bibnamefont{et~al.}, \bibinfo{journal}{Phys.Rept.}
  \textbf{\bibinfo{volume}{438}}, \bibinfo{pages}{1} (\bibinfo{year}{2007}),
  \eprint{hep-ph/0611040}.

\bibitem[{\citenamefont{Abel}(1996)}]{Abel:1996cr}
\bibinfo{author}{\bibfnamefont{S.~A.} \bibnamefont{Abel}},
  \bibinfo{journal}{Nucl. Phys.} \textbf{\bibinfo{volume}{B480}},
  \bibinfo{pages}{55} (\bibinfo{year}{1996}), \eprint{hep-ph/9609323}.

\bibitem[{\citenamefont{Dobrescu and Matchev}(2000)}]{Dobrescu:2000yn}
\bibinfo{author}{\bibfnamefont{B.~A.} \bibnamefont{Dobrescu}} \bibnamefont{and}
  \bibinfo{author}{\bibfnamefont{K.~T.} \bibnamefont{Matchev}},
  \bibinfo{journal}{JHEP} \textbf{\bibinfo{volume}{09}}, \bibinfo{pages}{031}
  (\bibinfo{year}{2000}), \eprint{hep-ph/0008192}.

\bibitem[{\citenamefont{Falkowski et~al.}(2010)\citenamefont{Falkowski, Krohn,
  Shelton, Thalapillil, and Wang}}]{Falkowski:2010hi}
\bibinfo{author}{\bibfnamefont{A.}~\bibnamefont{Falkowski}},
  \bibinfo{author}{\bibfnamefont{D.}~\bibnamefont{Krohn}},
  \bibinfo{author}{\bibfnamefont{J.}~\bibnamefont{Shelton}},
  \bibinfo{author}{\bibfnamefont{A.}~\bibnamefont{Thalapillil}},
  \bibnamefont{and} \bibinfo{author}{\bibfnamefont{L.-T.} \bibnamefont{Wang}}
  (\bibinfo{year}{2010}), \eprint{1006.1650}.

\bibitem[{\citenamefont{Chen et~al.}(2010)\citenamefont{Chen, Nojiri, and
  Sreethawong}}]{Chen:2010wk}
\bibinfo{author}{\bibfnamefont{C.-R.} \bibnamefont{Chen}},
  \bibinfo{author}{\bibfnamefont{M.~M.} \bibnamefont{Nojiri}},
  \bibnamefont{and}
  \bibinfo{author}{\bibfnamefont{W.}~\bibnamefont{Sreethawong}}
  (\bibinfo{year}{2010}), \eprint{1006.1151}.

\bibitem[{\citenamefont{Raby}(1997)}]{Raby:1997pb}
\bibinfo{author}{\bibfnamefont{S.}~\bibnamefont{Raby}},
  \bibinfo{journal}{Phys.Rev.} \textbf{\bibinfo{volume}{D56}},
  \bibinfo{pages}{2852} (\bibinfo{year}{1997}), \eprint{hep-ph/9702299}.

\bibitem[{\citenamefont{Baer et~al.}(1999)\citenamefont{Baer, Cheung, and
  Gunion}}]{Baer:1998pg}
\bibinfo{author}{\bibfnamefont{H.}~\bibnamefont{Baer}},
  \bibinfo{author}{\bibfnamefont{K.-m.} \bibnamefont{Cheung}},
  \bibnamefont{and} \bibinfo{author}{\bibfnamefont{J.~F.}
  \bibnamefont{Gunion}}, \bibinfo{journal}{Phys.Rev.}
  \textbf{\bibinfo{volume}{D59}}, \bibinfo{pages}{075002}
  (\bibinfo{year}{1999}), \eprint{hep-ph/9806361}.

\bibitem[{\citenamefont{Mafi and Raby}(2001)}]{Mafi:2000kg}
\bibinfo{author}{\bibfnamefont{A.}~\bibnamefont{Mafi}} \bibnamefont{and}
  \bibinfo{author}{\bibfnamefont{S.}~\bibnamefont{Raby}},
  \bibinfo{journal}{Phys.Rev.} \textbf{\bibinfo{volume}{D63}},
  \bibinfo{pages}{055010} (\bibinfo{year}{2001}), \eprint{hep-ph/0009202}.

\bibitem[{\citenamefont{Raby}(1998)}]{Raby:1997ba}
\bibinfo{author}{\bibfnamefont{S.}~\bibnamefont{Raby}},
  \bibinfo{journal}{Phys.Lett.} \textbf{\bibinfo{volume}{B422}},
  \bibinfo{pages}{158} (\bibinfo{year}{1998}), \eprint{hep-ph/9712254}.

\bibitem[{\citenamefont{Mafi and Raby}(2000)}]{Mafi:1999dg}
\bibinfo{author}{\bibfnamefont{A.}~\bibnamefont{Mafi}} \bibnamefont{and}
  \bibinfo{author}{\bibfnamefont{S.}~\bibnamefont{Raby}},
  \bibinfo{journal}{Phys.Rev.} \textbf{\bibinfo{volume}{D62}},
  \bibinfo{pages}{035003} (\bibinfo{year}{2000}), \eprint{hep-ph/9912436}.

\bibitem[{\citenamefont{Arkani-Hamed and
  Dimopoulos}(2005)}]{ArkaniHamed:2004fb}
\bibinfo{author}{\bibfnamefont{N.}~\bibnamefont{Arkani-Hamed}}
  \bibnamefont{and}
  \bibinfo{author}{\bibfnamefont{S.}~\bibnamefont{Dimopoulos}},
  \bibinfo{journal}{JHEP} \textbf{\bibinfo{volume}{06}}, \bibinfo{pages}{073}
  (\bibinfo{year}{2005}), \eprint{hep-th/0405159}.

\bibitem[{CMS()}]{CMSRHadrons}
\bibinfo{note}{{CMS Collaboration, CMS PAS EXO-10-004}}.

\bibitem[{\citenamefont{Arvanitaki et~al.}(2007)\citenamefont{Arvanitaki,
  Dimopoulos, Pierce, Rajendran, and Wacker}}]{latedecay}
\bibinfo{author}{\bibfnamefont{A.}~\bibnamefont{Arvanitaki}},
  \bibinfo{author}{\bibfnamefont{S.}~\bibnamefont{Dimopoulos}},
  \bibinfo{author}{\bibfnamefont{A.}~\bibnamefont{Pierce}},
  \bibinfo{author}{\bibfnamefont{S.}~\bibnamefont{Rajendran}},
  \bibnamefont{and} \bibinfo{author}{\bibfnamefont{J.~G.}
  \bibnamefont{Wacker}}, \bibinfo{journal}{Phys. Rev.}
  \textbf{\bibinfo{volume}{D76}}, \bibinfo{pages}{055007}
  (\bibinfo{year}{2007}), \eprint{hep-ph/0506242}.

\bibitem[{\citenamefont{Abazov et~al.}(2007)}]{Abazov:2007ht}
\bibinfo{author}{\bibfnamefont{V.}~\bibnamefont{Abazov}} \bibnamefont{et~al.}
  (\bibinfo{collaboration}{D0 Collaboration}),
  \bibinfo{journal}{Phys.Rev.Lett.} \textbf{\bibinfo{volume}{99}},
  \bibinfo{pages}{131801} (\bibinfo{year}{2007}), \eprint{arXiv:0705.0306}.

\bibitem[{Col()}]{Collaboration:2010uf}
\bibinfo{note}{{CMS Collaboration, arXiv:1011.5861}}.

\bibitem[{\citenamefont{Farrar et~al.}(2010)\citenamefont{Farrar, Mackeprang,
  Milstead, and Roberts}}]{Farrar:2010ps}
\bibinfo{author}{\bibfnamefont{G.~F.} \bibnamefont{Farrar}},
  \bibinfo{author}{\bibfnamefont{R.}~\bibnamefont{Mackeprang}},
  \bibinfo{author}{\bibfnamefont{D.}~\bibnamefont{Milstead}}, \bibnamefont{and}
  \bibinfo{author}{\bibfnamefont{J.~P.} \bibnamefont{Roberts}}
  (\bibinfo{year}{2010}), \eprint{1011.2964}.

\bibitem[{\citenamefont{Alwall et~al.}(2007)\citenamefont{Alwall, Demin,
  de~Visscher, Frederix, Herquet et~al.}}]{Madgraph}
\bibinfo{author}{\bibfnamefont{J.}~\bibnamefont{Alwall}},
  \bibinfo{author}{\bibfnamefont{P.}~\bibnamefont{Demin}},
  \bibinfo{author}{\bibfnamefont{S.}~\bibnamefont{de~Visscher}},
  \bibinfo{author}{\bibfnamefont{R.}~\bibnamefont{Frederix}},
  \bibinfo{author}{\bibfnamefont{M.}~\bibnamefont{Herquet}},
  \bibnamefont{et~al.}, \bibinfo{journal}{JHEP}
  \textbf{\bibinfo{volume}{0709}}, \bibinfo{pages}{028} (\bibinfo{year}{2007}),
  \eprint{0706.2334}.

\bibitem[{\citenamefont{Thomas and Wells}(1998)}]{ThomasWells}
\bibinfo{author}{\bibfnamefont{S.~D.} \bibnamefont{Thomas}} \bibnamefont{and}
  \bibinfo{author}{\bibfnamefont{J.~D.} \bibnamefont{Wells}},
  \bibinfo{journal}{Phys. Rev. Lett.} \textbf{\bibinfo{volume}{81}},
  \bibinfo{pages}{34} (\bibinfo{year}{1998}), \eprint{hep-ph/9804359}.

\bibitem[{\citenamefont{Buckley et~al.}(2009)\citenamefont{Buckley, Randall,
  and Shuve}}]{RandallBuckley}
\bibinfo{author}{\bibfnamefont{M.~R.} \bibnamefont{Buckley}},
  \bibinfo{author}{\bibfnamefont{L.}~\bibnamefont{Randall}}, \bibnamefont{and}
  \bibinfo{author}{\bibfnamefont{B.}~\bibnamefont{Shuve}}
  (\bibinfo{year}{2009}), \eprint{0909.4549}.

\end{thebibliography}
\bibliographystyle{apsrev}

\end{document}